# The structure of near stoichiometric Ge-Ga-Sb-S glasses: a reverse Monte Carlo study

I. Pethes[a,1], V. Nazabal[b], R. Chahal[b], B. Bureau[b], I. Kaban[c], B. Beuneu[d], J. Bednarcik[e], P. Jóvári[a]

[a]Wigner Research Centre for Physics, Hungarian Academy of Sciences, H-1525 Budapest, POB 49, Hungary

[b]Institut Sciences Chimiques de Rennes, UMR-CNRS 6226, Campus de Beaulieu, Université de Rennes 1, 35042 Rennes Cedex, France

[c]IFW Dresden, Institute for Complex Materials, Helmholtzstr. 20, 01069 Dresden, Germany

[d]Laboratoire Léon Brillouin, CEA-Saclay 91191 Gif sur Yvette Cedex France

[e]Deutsches Elektronen-Synchrotron − A Research Centre of the Helmholtz Association, Notkestraße 85, D-22607 Hamburg, Germany

Abstract

The structure of $Ge_{22}Ga_3Sb_{10}S_{65}$ and $Ge_{15}Ga_{10}Sb_{10}S_{65}$ glasses was investigated by neutron diffraction (ND), X-ray diffraction (XRD), and extended X-ray absorption fine structure (EXAFS) measurements at the Ge, Ga and Sb K-edges. Experimental data sets were fitted simultaneously in the framework of the reverse Monte Carlo (RMC) simulation technique. Short range order parameters were determined from the obtained large-scale configurations. It was found that the coordination numbers of Ge, Sb and S are around the values predicted by the Mott-rule (4, 3 and 2, respectively). The Ga atoms have on average 4 nearest neighbors. The structure of these stoichiometric glasses can be described by the chemically ordered network model: Ge-S, Ga-S and Sb-S bonds are the most important. Long Sb-S distances (0.3 – 0.4 Å higher than the usual covalent bond lengths) are observed, suggesting that Sb atoms can be found in various local environments.

*Keywords:* chalcogenide glasses, structure, RMC, diffraction, EXAFS, Ge-Ga-Sb-S

[1]Corresponding author. E-mail address: pethes.ildiko@wigner.mta.hu

## 1. Introduction

Chalcogenide glasses have remarkable physical properties such as their wide transparency window in the mid-infrared range far beyond that of the silica glasses (up to 12, 16, 28 microns for sulfide-, selenide- and telluride-based glasses respectively) [1], large linear and nonlinear refractive indices [2 - 5], low phonon energy [6, 7] or unique photosensitivity [8]. Passive optical properties of these glasses can be tuned by their chemical composition what makes them suitable for various applications in infrared (IR) optic, using these chalcogenide glasses as IR lenses, optical fibers or optical waveguides [9 - 15]. Their utilization can be further expanded by rare-earth doping which makes them applicable as active optical media. They are or may be used for optical fiber amplifiers [16], sensors and detectors [17 - 19] or as laser devices [7, 20, 21]

Germanium based sulfide glasses, such as Ge-Ga-S, have a relatively good rare-earth ion solubility thanks to the addition of gallium in the glass network [22 - 24]. Moreover, the Ge-Ga-S system can be stabilized against crystallization by the addition of arsenic or antimony [10, 25, 26] and efficient rare earth ion solubility to enable fabrication of conventional and tapered fibers, thin films and rib waveguides presenting an efficient luminescence [27 - 35].

A comprehensive description of the structure of these glasses can be helpful for the better understanding of their properties. The structure of the chalcogenide glasses can be described as a covalently connected network of the participant elements. In glasses consisting elements from the 14-15-16 groups of the periodic table, the total coordination number of the elements ($N_i$) follows the Mott-rule [36]: It is equal to 8-$N$, where $N$ is the number of electrons in the valence shell of the $i$th element (e. g. Ge-As-Se [37], Ge-As-Te [38], Ge-Sb-Se [39], Ge-Sb-Te [40]). However glasses containing group 13 elements can deviate from this rule, and the total coordination number of Ga or In can be four instead of three [41 - 44].

The structure of several chalcogenide glasses can be described in the framework of the chemically ordered network model (CONM) [45, 46]. This model predicts that M-Ch bonds are preferable, where M denotes the elements from groups 13, 14 or 15, and Ch means the chalcogen element. The structure of the stoichiometric glass can be built from tetrahedral and/or pyramidal units such as [GeCh$_{4/2}$] or [SbCh$_{3/2}$]. The M-M and Ch-Ch bonds are present only in non-stoichiometric glasses: M-M bonds in Ch-deficit (Ch-under stoichiometric) compositions and Ch-Ch bonds in Ch-rich (Ch-over stoichiometric) glasses.

There are several publications about the structure of the ternary Ge-Ga-S glasses [43, 47 - 51]. It was demonstrated that Ge and S atoms follow the Mott-rule and their average coordination numbers are 4 and 2 respectively, but the average coordination number of Ga is higher than the predicted 3. $[GeS_{4/2}]$, $[GaS_{4/2}]$ units were reported in these glasses, as well as $[S_3Ge(Ga)-Ge(Ga)S_3]$ ethane-like units in S-deficient samples. A small amount of M-M and S-S bonds in stoichiometric samples showed some violation in the chemical order [47, 49].

In Ge-Sb-S glasses $[GeS_{4/2}]$ and $[SbS_{3/2}]$ units are the main building blocks as shown in a lot of studies [52 - 57]. S-S bonds were found in S-rich compositions [52, 55, 57]. M-M bonds were reported in S-deficient samples [52, 57, 58] and in stoichiometric compositions as well [53, 55], which shows some chemical disorder.

The structure of quaternary Ge-Ga-Sb-S glasses is much less known. Raman scattering spectroscopy and mass-spectrometry investigations [26, 30, 59 - 61] suggested that the main building blocks are $[GeS_{4/2}]$, $[GaS_{4/2}]$ and $[SbS_{3/2}]$ units in these glasses too. But there is no comprehensive study on their structure on the atomic level.

In this paper we present our X-ray diffraction (XRD), neutron diffraction (ND) and extended X-ray absorption fine structure (EXAFS) measurements on $Ge_{22}Ga_3Sb_{10}S_{65}$ and $Ge_{15}Ga_{10}Sb_{10}S_{65}$ glasses. Short range order parameters are investigated by combining the experimental results using the reverse Monte Carlo (RMC) simulation technique [62, 63].

2. Experimental

Samples were prepared by conventional melt quenching. High purity raw materials were used for preparing $Ge_{22}Ga_3Sb_{10}S_{65}$ and $Ge_{15}Ga_{10}Sb_{10}S_{65}$, i.e. 5N for germanium, gallium, antimony or sulphur. Commercial sulphur was further purified by successive distillations to remove carbon ($CO_2$, $CS_2$, COS) and hydrates or sulphide hydride ($H_2O$, OH, SH) impurities. Then, the required amounts of chemical reagents were put in silica ampoules and pumped under vacuum (~$10^{-4}$ mbar) for a few hours. The tubes were then sealed and heated at 850°C for 12h in a rocking furnace to ensure the homogenization of the melt. After water quenching, the glass rods were annealed near their glass transition temperatures for 6h. The densities of the samples were determined using a Mettler Toledo XS64 system measuring the weights of the samples in air and water. The density values are shown in Table 1.

Neutron diffraction experiments were carried out at the 7C2 diffractometer of LLB (Saclay, France).

The wavelength of incident radiation was 0.72 Å. Powdered samples were filled into thin walled (0.1 mm) vanadium containers of 6 mm diameter. Raw data were corrected for background scattering, detector efficiency and multiple scattering.

The XRD structure factors were measured at the beamline P07 of the Petra III source (Hamburg, Germany). The energy of the incident radiation was 80 keV. Raw intensities were recorded by a Perkin-Elmer 2D detector. Wavelength, detector position and tilting were determined by measuring a $LaB_6$ standard. 2D counts were integrated circularly and corrected for background, absorption, detector solid angle and Compton scattering by the program Fit2D [64]. Corrected ND and XRD structure factors are presented in Fig. 1.

The extended X-ray absorption fine structure measurements at the Ge, Ga and Sb K-edges were carried out at HASYLAB (beamline X). Glassy samples were finely ground, mixed with cellulose and pressed into tablets. The sample quantities in the tablets were adjusted to the compositions and to the selected edges. The spectra were collected in the transmission mode using fixed exit double-crystal Si(111). The intensities before and after the sample as well as after the reference samples were recorded by ionization chambers filled with a mixture of Ar/Kr (~10% absorption), Ar (~50% absorption) and Kr (~100% absorption), respectively. Intensities were converted to $\chi(k)$ curves by the Viper programme [65]. Filtered EXAFS curves are plotted in Fig. 2. The $\chi(k)$ curves are multiplied by $k^3$ to emphasize high-$k$ oscillations decaying quickly with $k$.

Both diffraction and EXAFS are sensitive to pair correlations. However, the ways the partial pair correlation functions are transformed to structure factors or $\chi(k)$ curves are different (see below). For this reason the most straightforward and economic procedure of creating models compatible with diffraction and EXAFS data simultaneously is to calculate first partial pair correlation functions and then refine them gradually to reproduce all measurements. In practice this goal is reached by the reverse Monte Carlo simulation technique.

## 3. Reverse Monte Carlo simulations

The reverse Monte Carlo method [62] is a powerful tool to get large three dimensional atomic configurations consistent with diffraction (ND or XRD) or EXAFS data. The experimental data sets can be fitted simultaneously by this simulation technique. Further physical and chemical properties, such as density or preferred bond angles and coordination numbers can be taken into account. During the simulations particles are moved around to minimize the discrepancies between the experimental and model curves, and finally particle configurations can be obtained which are compatible with all the experimental data sets within their experimental error. From these configurations short range order parameters (partial pair correlation functions, nearest neighbor distances, average coordination numbers etc.) can be determined.

The $S_{mod}(Q)$ model structure factor can be calculated from the partial pair correlation functions ($g_{ij}(r)$) according to the following equations:

$$S_{ij}(Q) - 1 = \frac{4\pi\rho_0}{Q} \int_0^\infty r(g_{ij}(r) - 1)\sin(Qr)dr \qquad (1)$$

$$S_{mod}(Q) = \sum_{i \leq j} w_{ij}^{N,X}(Q) S_{ij}(Q). \qquad (2)$$

Here $Q$ is the scattering variable, $\rho_0$ is the average number density. The $w_{ij}^{N,X}$ neutron and X-ray scattering weights are given by Eqs. (3-5):

$$w_{ij}^N = (2 - \delta_{ij}) \frac{c_i c_j b_i b_j}{\langle f(Q) \rangle^2}, \qquad (3)$$

$$w_{ij}^X(Q) = (2 - \delta_{ij}) \frac{c_i c_j f_i(Q) f_j(Q)}{\langle f(Q) \rangle^2} \qquad (4)$$

and

$$\langle f(Q) \rangle^2 = [\sum_i c_i f_i(Q)]^2 \qquad (5)$$

where $\delta_{ij}$ is the Kronecker delta, $b_i$ is the coherent neutron scattering length, $c_i$ is the atomic concentration, and $f_i(Q)$ is the atomic form factor.

The model EXAFS curves are calculated according to Eq. (6)

$$\chi_i^{mod}(k) = \sum_j 4\pi\rho_0 c_j \int_0^R r^2 \gamma_{ij}(k,r) g_{ij}(r) dr, \qquad (6)$$

where i is the index of the absorbing component, and $\gamma_{ij}(k,r)$ is the photoelectron backscattering matrix.

In this study the two total structure factors from neutron and X-ray diffraction measurements and 3 $\chi(k)$ EXAFS (Ge-, Ga- and Sb- K-edges) data sets are fitted by the RMC++ code [63]. The EXAFS backscattering coefficients were calculated by the FEFF8.4 program [66]. The $w_{ij}^N$ neutron weighting factors of the investigated glasses are shown in Table 2, while the $Q$-dependent X-ray weights are presented in Fig. 3.

The cubic simulation boxes contained 8000 atoms in test runs and 30 000 particles in the final runs presented here. The box sizes were determined according to the experimental densities, which are shown in Table 1. Several test runs were performed to determine the various bond types, which are necessary to get proper fit for the experimental data sets. Hence different minimum interatomic distances (cut-offs) were used depending on whether the specific bond was allowed or not. The applied cut-off distances are presented in Table 3. Initial particle configurations were created by placing the atoms randomly in the cubic box and moving them around to satisfy the cut-off distance requirements. Initial sigma parameters used to calculate the RMC cost function [62, 67] were reduced in three steps and have final values as 0.001-0.0015 for XRD and ND, and $5 \times 10^{-6} - 2 \times 10^{-5}$ for EXAFS data sets. The number of accepted moves was around $1\text{-}2 \times 10^7$ steps.

Some 'background' coordination constrains were used to avoid segregated atoms and unrealistically high (7 or more for Ge and Ga, 6 or more for Sb and 4 or more for S atoms) and low (1 and 2 for Ge and 1 for Sb and Ga) coordination numbers. (Models with only these 'background' constraints will be called 'free' or 'unconstrained' hereafter.)

The tested models were compared by checking their *R*-factors, which measures the quality of the fit:

$$R = \frac{\sqrt{\sum_i \left(S_{\text{mod}}(Q_i) - S_{\text{exp}}(Q_i)\right)^2}}{\sqrt{\sum_i S_{\text{exp}}^2(Q_i)}} \tag{7}$$

Here $S_{\text{mod}}(Q_i)$ and $S_{\text{exp}}(Q_i)$ are the model and experimental structure factors and the summation runs over the experimental $Q_i$ values. The *R*-factor for EXAFS data is calculated similarly, using the corresponding model and experimental $\chi(k)$ curves. The relative *R*-factor of a configuration (with respect to a reference model) is defined as the ratio of the *R*-factor of the configuration and that of the reference model. A simple average of the relative *R*-factors of the five experimental data sets gives a cumulative relative *R*-factor ($R_c$), which was used to categorize the investigated model.

A four component system has 10 independent partial pair correlation functions (PPCF, $g_{ij}(r)$), meaning that the same number of independent measurements would be necessary to determine the PPCFs in a purely algebraic way. Various constraints (coordination constraints, forbidding some bonds etc.) can help to decrease the uncertainty of the short range order parameters of the system. For this reason it is advisable to keep the number of allowed bonds at its minimum value. The M-S bonds (Ge-S, Ga-S and Sb-S) were always allowed, the necessity of the M-M and S-S bonds were investigated.

The average partial coordination numbers ($N_{ij}$) were calculated by integrating the partial pair correlation functions up to the first minimum ($r_{min}$) :

$$N_{ij} = 4\pi r^2 \rho_0 c_j \int_0^{r_{min}} g_{ij}(r) dr \tag{8}$$

where $\rho_0$ is the average density, $c_j$ is the concentration of the *j*th element. To estimate the uncertainty of the average coordination numbers dedicated simulation runs were performed. The value of the investigated $N_{ij}$ was forced to change systematically (in ± 10% increments). By monitoring the *R*-factors, the range of $N_{ij}$ values, in which the quality of the fit is appropriate, can be determined.

The $N_i$ total coordination numbers are calculated as:

$$N_i = \sum_j N_{ij}. \tag{9}$$

## 4. Results and discussion

Experimental total structure factors *S(Q)* and $k^3$ weighted, filtered EXAFS signals (*χ(k)* curves) are presented in Figs. 1 and 2. Model configurations were obtained by simultaneous fitting the 5 experimental data sets for both compositions.

Assuming that the Ge, Ga, Sb and S atoms are 4-, 4-, 3- and 2-fold coordinated respectively, the investigated compositions are stoichiometric. According to the CONM and earlier Raman scattering spectroscopy results [26, 30, 61] in stoichiometric glasses M-S (M=Ge, Ga, Sb) bonds are the most dominant, and only a small amount of S-S and M-M bonds is expected. Several test runs were carried out to determine the bond types which are significant in the studied glasses. It was found that applying the model in which only Ge-S, Ga-S and Sb-S bonds are allowed, the experimental data sets can be fitted properly. The simulated curves obtained by applying this model are also shown in Figs. 1 and 2. The presence of other bond types did not improve the quality of the fits. It means that neither of the Ge-Ge, Ge-Ga, Ge-Sb, Ga-Ga, Ga-Sb, Sb-Sb and S-S coordination numbers are higher than 0.3-0.4, which is the sensitivity of our method.

The partial pair correlation functions calculated from the obtained particle configurations are presented in Fig. 4. The Ge-S, Ga-S and Sb-S PPCFs have sharp peaks in the 2.1-3.0 Å region, which is in the first coordination shell of the participant atoms. Positions of the first peaks are shown in Table 4. The average nearest neighbor distances are independent of the compositions within their errors.

The first peak of the Ge-S and Ga-S PPCFs are around 2.22 Å and 2.27-2.28 Å, respectively. These values agree well with that obtained earlier in ternary Ge-Ga-S glasses by XAFS spectroscopy [48] (2.22 Å and 2.28 Å) and by RMC method with EXAFS and XRD measurements [43].

The Sb-S bond length (2.44 Å) is somewhat shorter than that found in Ge-Sb-S glasses by ND (2.48 Å [56]), but agrees with the values obtained by EXAFS for Sb-S systems (2.45-2.46 Å [68]) and by RMC method with EXAFS, ND and XRD measurements for ternary Ge-Sb-S glasses (2.45 Å [57]). (The Sb-S bond length of Kakinuma et al. [56] is deduced from the total scattering function of S-deficient Ge-Sb-S glasses, where the presence of Ge-Sb pairs (with bond length around 2.61 -2.65 Å [57, 58]) can cause the shift of the peak to higher values.

The average coordination numbers are calculated up to 2.9 Å for $N_{GeS}$ and $N_{GaS}$ and to 3.0 Å for $N_{SbS}$; the values are presented in Table 5. In these unconstrained simulation runs only the above mentioned 'background' coordination constraints were used, the average coordination numbers of the different pairs were allowed to change freely. The average coordination number of sulphur, antimony and germanium are around 2, 3 and 4 respectively, as they were expected.

The average coordination number of gallium ($N_{Ga}$) is definitely higher than 3 for both compositions. Due to the small amount of gallium in the $Ge_{22}Ga_3Sb_{10}S_{65}$ glass, the uncertainty of this number is inevitably higher. Simulation runs were carried out to test the coordination number of Ga. In these runs the average coordination number of the Ga-S pair ($N_{GaS}$, which is equal to $N_{Ga}$ since Ga-Ge, Ga-Ga and Ga-Sb bonds were forbidden) was constrained at different values and the changes in the quality of the fits were monitored. The *R*-factors of the curves were an average 20 % higher for those test runs in which the Ga-S coordination number was constrained to be 3, and they only slightly increased (below 5%) when the $N_{GaS}$= 4 constraint was used. It was thus concluded that the total coordination number of Ga must be close to 4. It is to be mentioned that *unconstrained* simulation runs gave 3.85 and 3.67 for $N_{Ga}$ in Ge-Ga-S glasses [43].

In all M-S curves smaller and less sharp second peaks can be observed in the 2.6 - 3.0 Å region (see Fig. 4). The significance of these peaks was tested with simulation runs in which the presence of these 'long bonds' were forbidden (by applying zero coordination constraints for $N_{GeS}$, $N_{GaS}$ and $N_{SbS}$ in the above mentioned region). It was found that the quality of the fits decreased for this model significantly (on the average the *R*-factors increased with 50%). The ND and Ge EXAFS fits of $Ge_{22}Ga_3Sb_{10}S_{65}$ glass with and without 'long bonds' are shown in Fig. 5. Further investigations showed that introducing M-M type bonds in the RMC model does not improve fit quality. It is more sensitive to Ge-S and Sb-S 'long bonds', which are equally important to get proper fits for the present set of experimental data. We also checked whether the finite range of diffraction data has any effect on the position of the 'long bonds'. Test runs with $Q_{max}$= 10 Å$^{-1}$ for ND and $Q_{max}$= 15.6 Å$^{-1}$ for XRD gave very similar PPCFs proving that the finite data range does not influence the presence of 'long bonds'.

Similar pair distances in the 2.6 - 3.0 Å region were reported earlier in the Ge-Sb-S system [56,58]. A 'hump' observed around 2.8 – 2.9 Å in the total XRD pair correlation function of S-deficient Ge-Sb-S glasses was explained by the presence of Sb-Sb pairs and/or by the distance of Ge-Ge pairs in edge-shared GeS$_{4/2}$ tetrahedra (Fig. 3 and 4 of Ref. [58]). A small peak can be seen in the $g(r)$ function of S-deficient Ge-Sb-S glasses, obtained from ND experiments (Fig. 3. of Ref. [56]) as well. Recently, S-rich, stoichiometric and S-deficient Ge-Sb-S glasses were investigated by RMC method fitting ND, XRD, Ge and Sb EXAFS data [57]. The presence of long Sb-S bonds was reported both in sulphur rich and poor region. Long Sb-S bonds are in crystalline Sb$_2$S$_3$, where SbS$_3$ and SbS$_5$ units can be found with 5 different Sb-S bond lengths in the 2.46 – 2.85 Å region [69, 70]. The combination of these units was proposed in glassy Sb$_2$S$_3$-As$_2$S$_3$-Tl$_2$S systems as well [71].

The above results suggest that the observed 'long bonds' are most probably of Sb-S type. The data analyzed in the present study allows the presence of long Ge-S bonds as well. However, as they were not found in Ge-Ga-S glasses [43] the presence of longer Ge-S type bonds is probably due to the 'cross talk' between Ge-S and Sb-S correlations. Further studies are under way to reveal where and why these long bonds appear in Sb$_2$S$_3$ based glasses.

The presence of Ge(Ga)-Ge(Ga) second neighbors that share two common S atoms (edge-shared Ge(Ga)S$_4$ tetrahedra) was also studied. The second coordination shells of Ge and Ga atoms are clearly visible in the corresponding PPCFs, as definite peaks around 3.5 Å. Analysis of the obtained configurations revealed, that the Ge(Ga)-Ge(Ga) distances in the 3.1 – 4.1 Å region originate not just from corner sharing (CS) pairs, but the typical distance of edge sharing (ES) Ge(Ga)-Ge(Ga) pairs is also around 3.1 – 3.6 Å (see Fig. 6). (Similar distances were also found in ternary Ge-Ga-S glasses in Ref. [43].) The ratio of Ge atoms participating in ES units is around 40% in the investigated glasses; the presence of ES units is supported by Raman spectroscopy.

The first sharp diffraction peak (FSDP) is considered as a sign of medium range order. According to Fig. 1 both X-ray and neutron diffraction structure factors exhibit a pronounced FSDP at about 1.07 Å$^{-1}$. A quick view at the $S_{ij}(Q)$ partial structure factors (Fig. 7) reveals that in case of the Ge-Ga-Sb-S glasses investigated the FSDP can be assigned mostly to Ge(Ga)-Ge(Ga) and Ge(Ga)-S correlations. With some rearrangement of Eq. (1) one gets the following:

$$S_{ij}(Q) - 1 = 4\pi\rho_0 \int_0^\infty r^2 (g_{ij}(r) - 1) \frac{\sin(Qr)}{Qr} dr \qquad (10)$$

Eq. (10) shows that an elemental component of $r^2(g_{ij}(r)-1)$ is transformed into a $\sin(x)/x$ type damped oscillation in $Q$-space. (The partial structure factor is then obtained as the superposition of such elemental oscillations.) The first maximum of $\sin(x)/x$ (after $x = 0$) is at $x \approx 7.72$. Thus, the correlation length corresponding to $Q_{max}$, the FSDP peak position is defined as $r_{FSDP} = 7.72/Q_{max}$ (see also the Ehrenfest relation [72]). In our case $r_{FSDP} \approx 7.2$ Å. This value is roughly equal to the double of the mean distance of centers of CS/ES tetrahedra.

## Conclusions

Short range order of $Ge_{22}Ga_3Sb_{10}S_{65}$ and $Ge_{15}Ga_{10}Sb_{10}S_{65}$ glasses was studied by neutron- and X-ray diffraction, combined with EXAFS measurements. The experimental data sets were fitted simultaneously by reverse Monte Carlo simulation technique. It has been established that Ge and Ga are both fourfold coordinated, while Sb and S atoms have 3 and 2 nearest neighbors, respectively. The structure of the examined glasses can be described by the chemically ordered network model: Ge-S, Ga-S and Sb-S bonds are the most preferred. The typical bond lengths are 2.22 Å for Ge-S, 2.28 Å for Ga-S and 2.44 Å for Sb-S pairs, but some longer distances, most probably Sb-S pairs, in the 2.6 – 3 Å region were also observed.

## Acknowledgments

The authors are thankful to ADEME, the French Environment and Energy Management Agency for financial support through COPTIK and Optigas projects (Grant No. ANR-15-CE39-0007). The neutron diffraction experiment was carried out at the ORPHÉE reactor, Laboratoire Léon Brillouin, CEA-Saclay, France. I. P. and P. J. are grateful to the National Research, Development and Innovation Office (NKFIH) of Hungary for financial support through Grant No. SNN 116198.

**Table 1** Densities of the investigated Ge-Ga-Sb-S glasses

| Composition | Density (g/cm$^3$) (± 0.01) | Number density (Å$^{-3}$) |
| --- | --- | --- |
| $Ge_{22}Ga_3Sb_{10}S_{65}$ | 3.15 | 0.03711 |
| $Ge_{15}Ga_{10}Sb_{10}S_{65}$ | 3.32 | 0.03930 |

**Table 2** Neutron scattering weights ($w_{ij}$) of the investigated glasses used for the calculation of the ND total structure factors.

| $i$-$j$ pairs | $Ge_{22}Ga_3Sb_{10}S_{65}$ | $Ge_{15}Ga_{10}Sb_{10}S_{65}$ |
| --- | --- | --- |
| Ge-Ge | 0.16546 | 0.07914 |
| Ge-Ga | 0.04018 | 0.09396 |
| Ge-Sb | 0.10236 | 0.07181 |
| Ge-S | 0.34008 | 0.23859 |
| Ga-Ga | 0.00244 | 0.02789 |
| Ga-Sb | 0.01243 | 0.04263 |
| Ga-S | 0.04129 | 0.14163 |
| Sb-Sb | 0.01583 | 0.01629 |
| Sb-S | 0.10519 | 0.10824 |
| S-S | 0.17474 | 0.17981 |

**Table 3** Minimum interatomic distances (cut-offs) applied in the simulation of Ge-Ga-Sb-S glasses (in Å)

| Pair | Bond is allowed | Bond is forbidden |
| --- | --- | --- |
| Ge-Ge | 2.25 | 2.75 |
| Ge-Ga | 2.35 | 2.75 |
| Ge-Sb | 2.35 | 2.9 |
| Ge-S | 2.0 | - |
| Ga-Ga | 2.45 | 2.9 |
| Ga-Sb | 2.5 | 2.9 |
| Ga-S | 2.05 | - |
| Sb-Sb | 2.7 | 3.15 |
| Sb-S | 2.25 | - |
| S-S | 1.95 | 2.9 |

**Table 4** Nearest neighbor distances (in Å) in the studied Ge-Ga-Sb-S glasses. The uncertainty of distances is usually ± 0.02 Å.

| | Glass composition | |
| --- | --- | --- |
| Pair | $Ge_{22}Ga_3Sb_{10}S_{65}$ | $Ge_{15}Ga_{10}Sb_{10}S_{65}$ |
| Ge-S | 2.22 | 2.22 |
| Ga-S | 2.27 | 2.28 |
| Sb-S | 2.44 | 2.44 |

**Table 5** Coordination numbers of the investigated compositions obtained for the model in which only Ge-S, Ga-S and Sb-S bonds are allowed and coordination constraints were not used.

|  | Glass composition | |
| --- | --- | --- |
|  | $Ge_{22}Ga_3Sb_{10}S_{65}$ | $Ge_{15}Ga_{10}Sb_{10}S_{65}$ |
| $N_{GeS} = N_{Ge}$ | 4.1 (-0.1 +0.3) | 4.4 (-0.2 +0.6) |
| $N_{GaS} = N_{Ga}$ | 3.8 (±0.4) | 4.1 (±0.4) |
| $N_{SbS} = N_{Sb}$ | 3.0 (-0.2 +0.3) | 3.1 (±0.3) |
| $N_{SGe}$ | 1.4 (-0.05 +0.1) | 1.03 (-0.06 +0.12) |
| $N_{SGa}$ | 0.175 (±0.015) | 0.63 (±0.06) |
| $N_{SSb}$ | 0.46 (-0.03 +0.05) | 0.48 (±0.05) |
| $N_S$ | 2.04 | 2.14 |

Figures

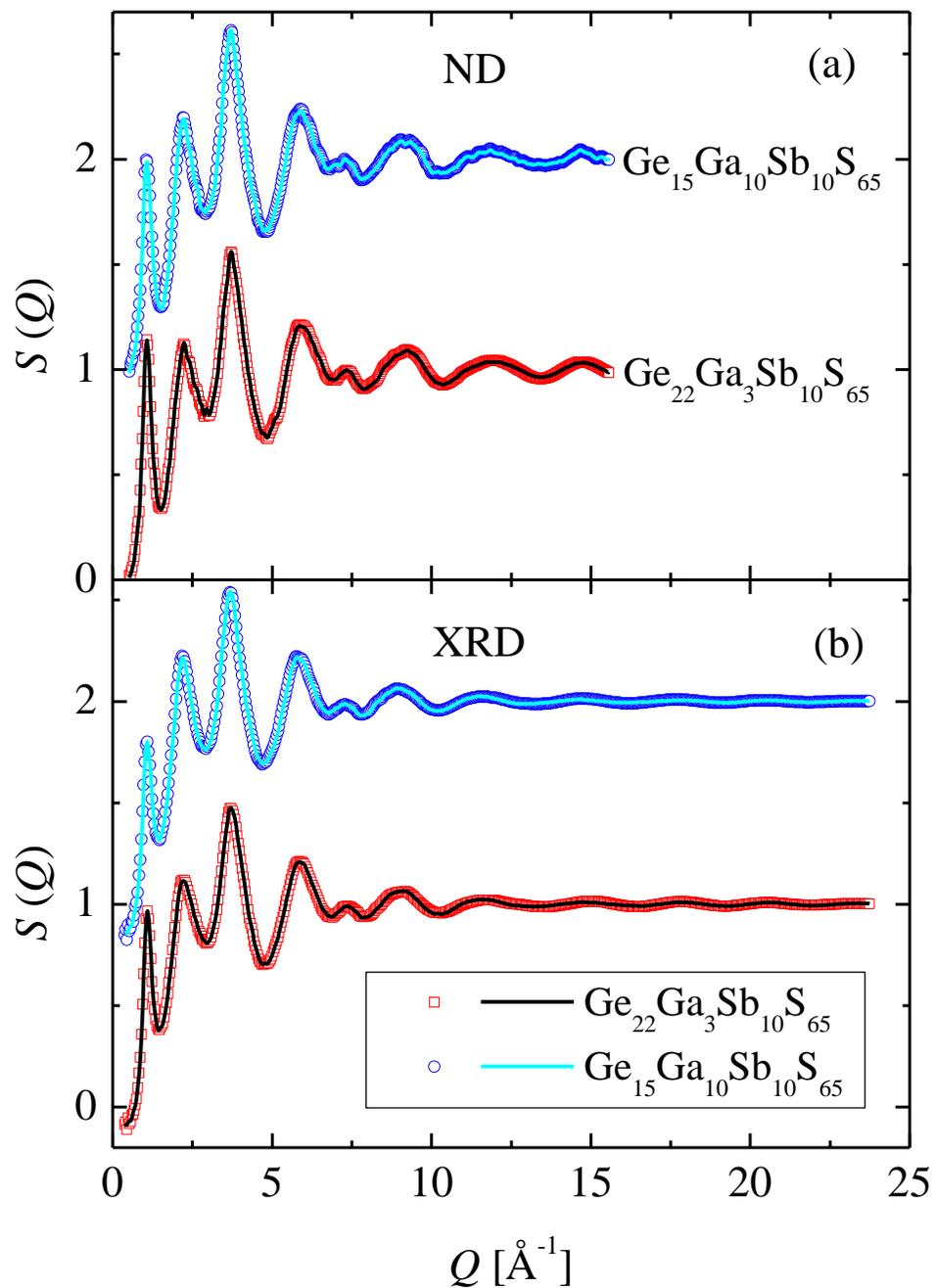

**Figure 1.** (a) ND and (b) XRD total structure factors of $Ge_{22}Ga_3Sb_{10}S_{65}$ and $Ge_{15}Ga_{10}Sb_{10}S_{65}$ (symbols) and fits (lines) obtained by RMC simulations. (The curves of the $Ge_{15}Ga_{10}Sb_{10}S_{65}$ glass are shifted upward for clarity.)

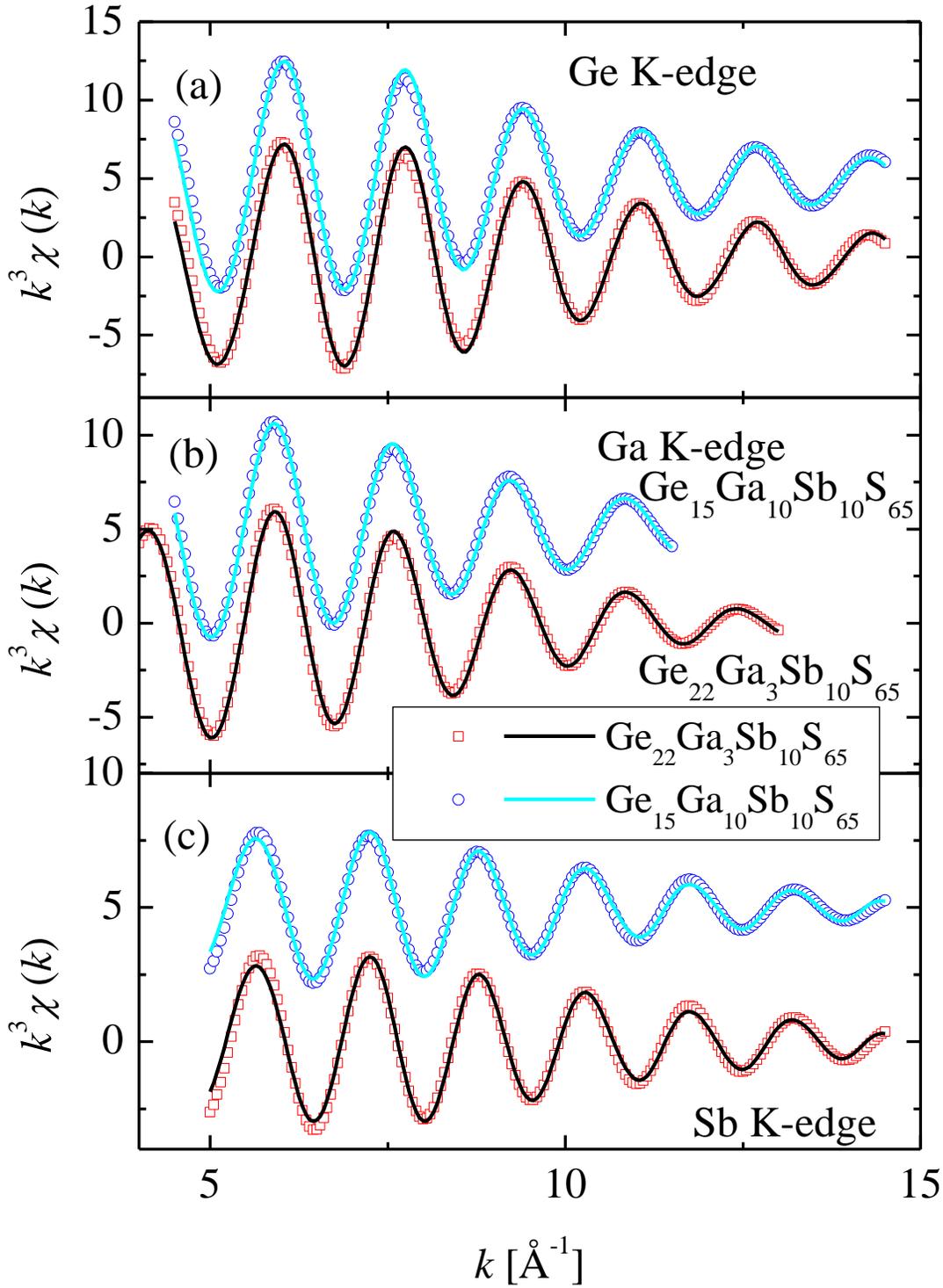

**Figure 2.** $k^3$ weighted, filtered (a) Ge, (b) Ga and (c) Sb K-edge EXAFS spectra of the investigated glasses (symbols) and fits (lines) obtained by RMC simulations. (The curves of the $Ge_{15}Ga_{10}Sb_{10}S_{65}$ glass are shifted upward for clarity.)

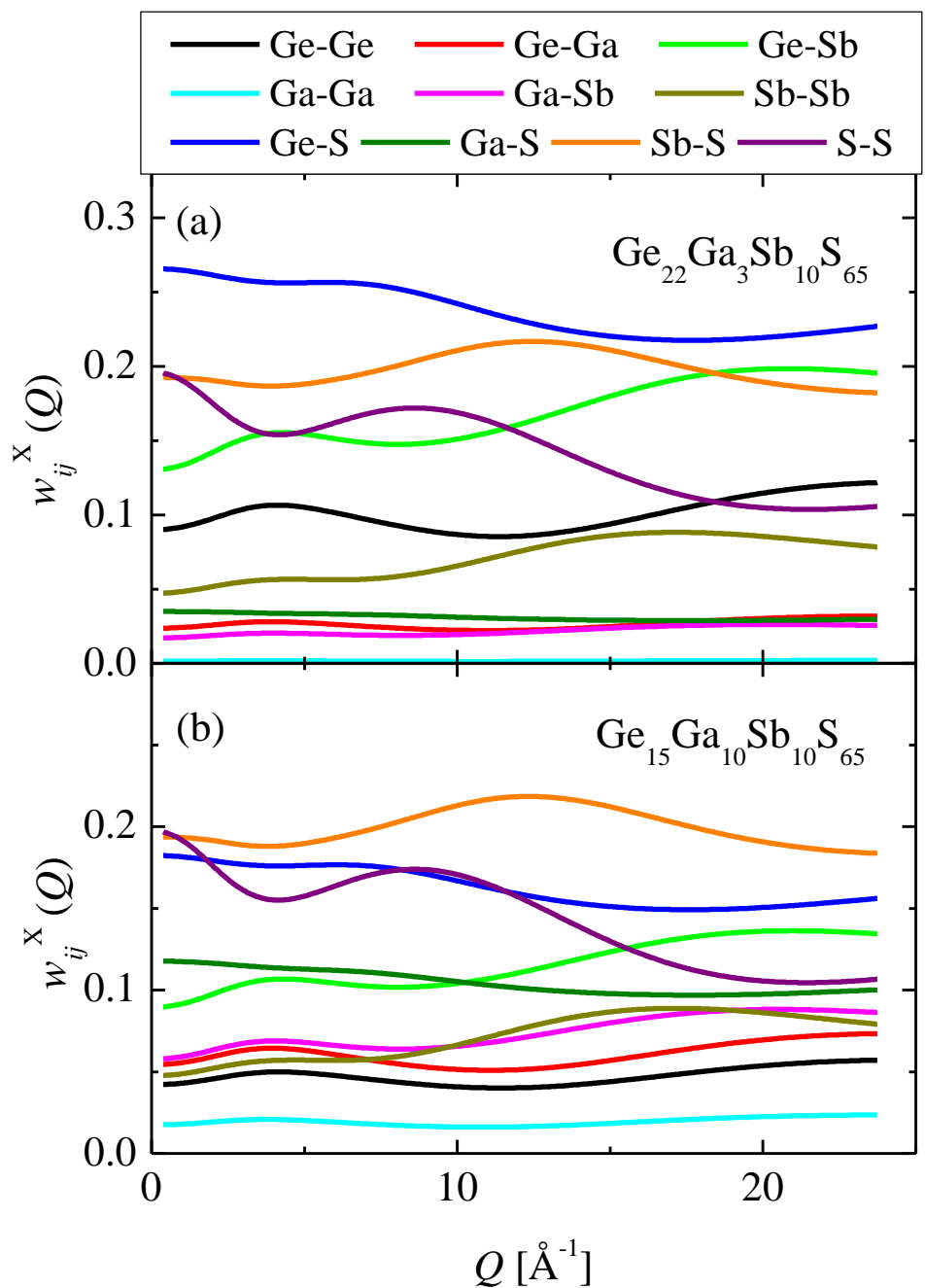

**Figure 3** X-ray diffraction weights of the partial structure factors of (a) $Ge_{22}Ga_3Sb_{10}S_{65}$ and (b) $Ge_{15}Ga_{10}Sb_{10}S_{65}$ glasses.

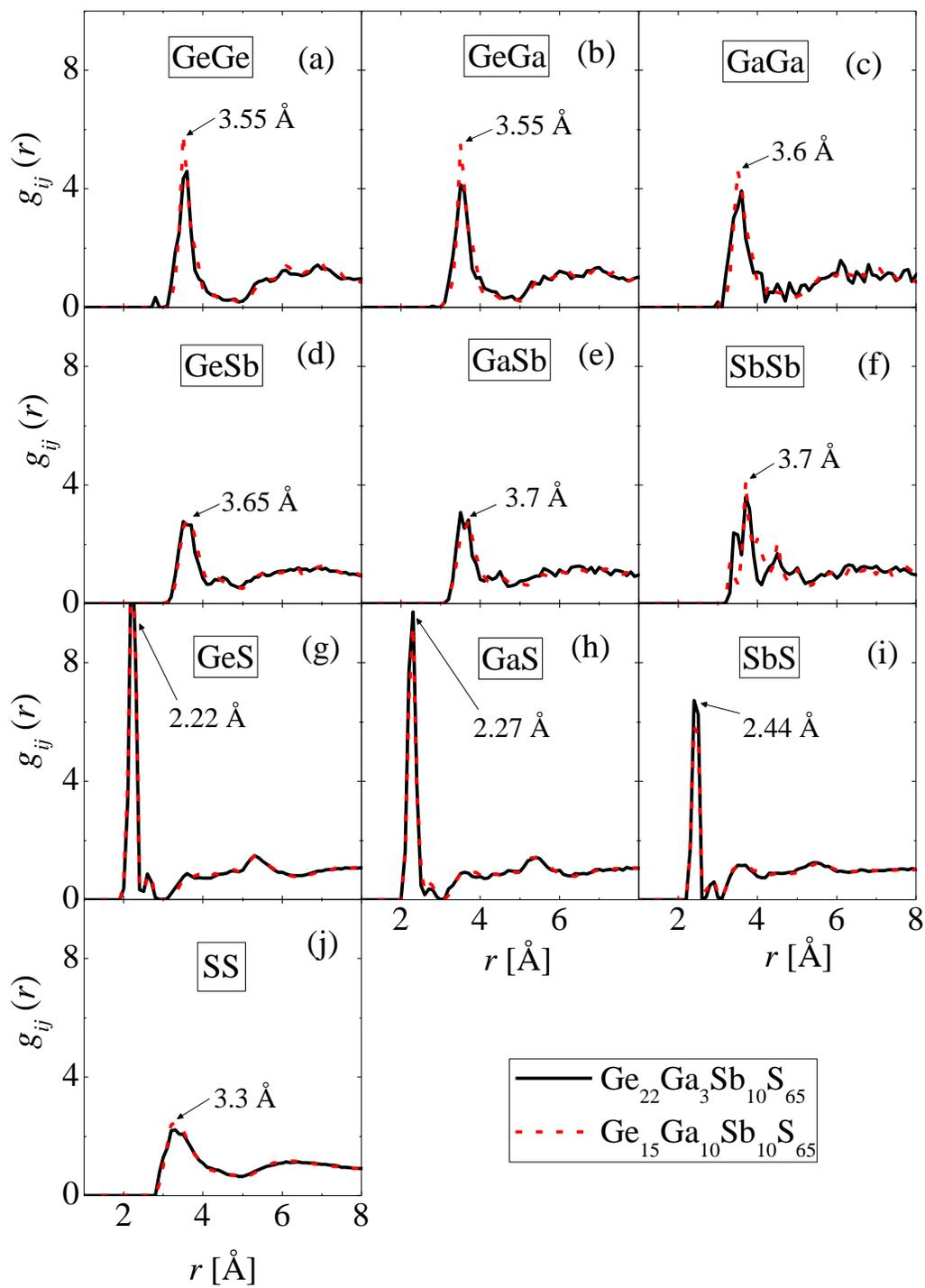

**Figure 4.** Partial pair correlation functions of the investigated Ge-Ga-Sb-S glasses.

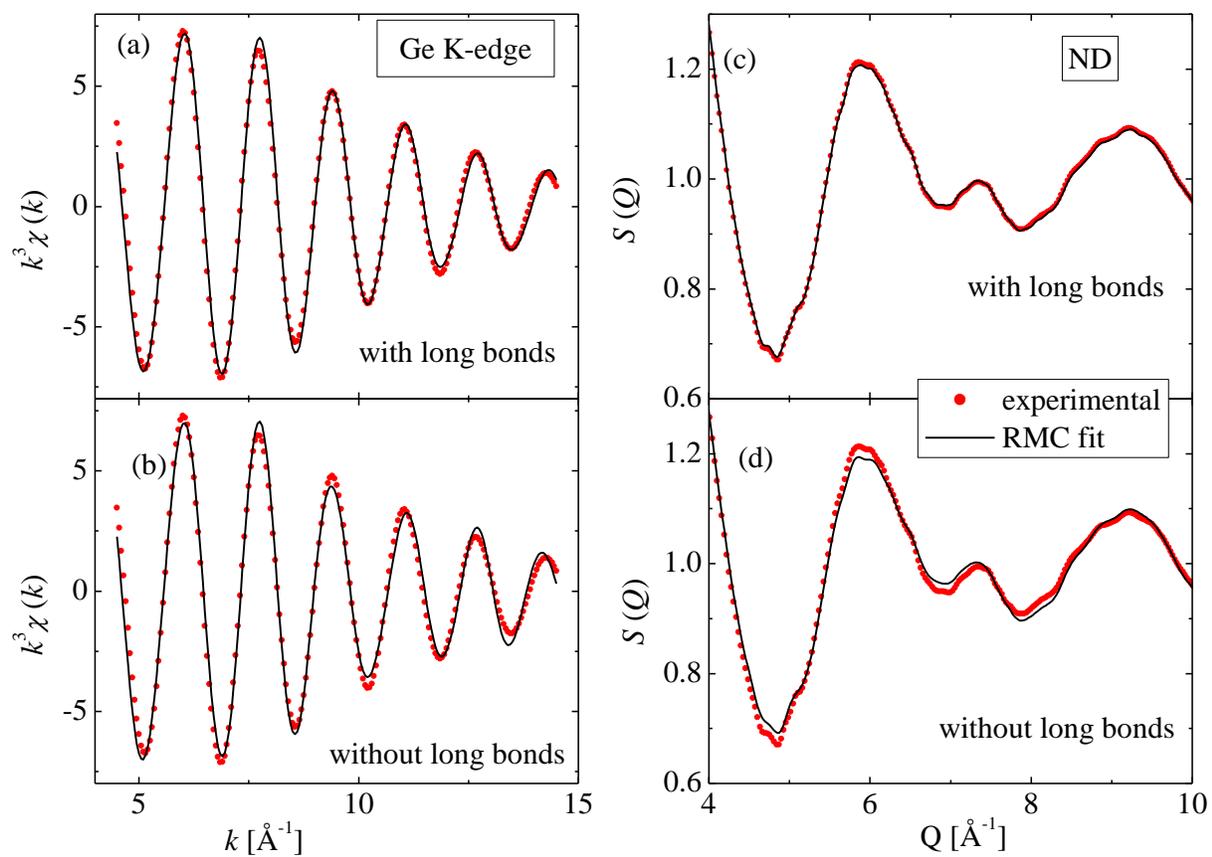

**Figure 5.** (a and b) Ge K-edge EXAFS and (c and d) ND fits of $Ge_{22}Ga_3Sb_{10}S_{65}$ glass (a and c) with and (b and d) without long Ge-S, Ga-S and Sb-S pairs.

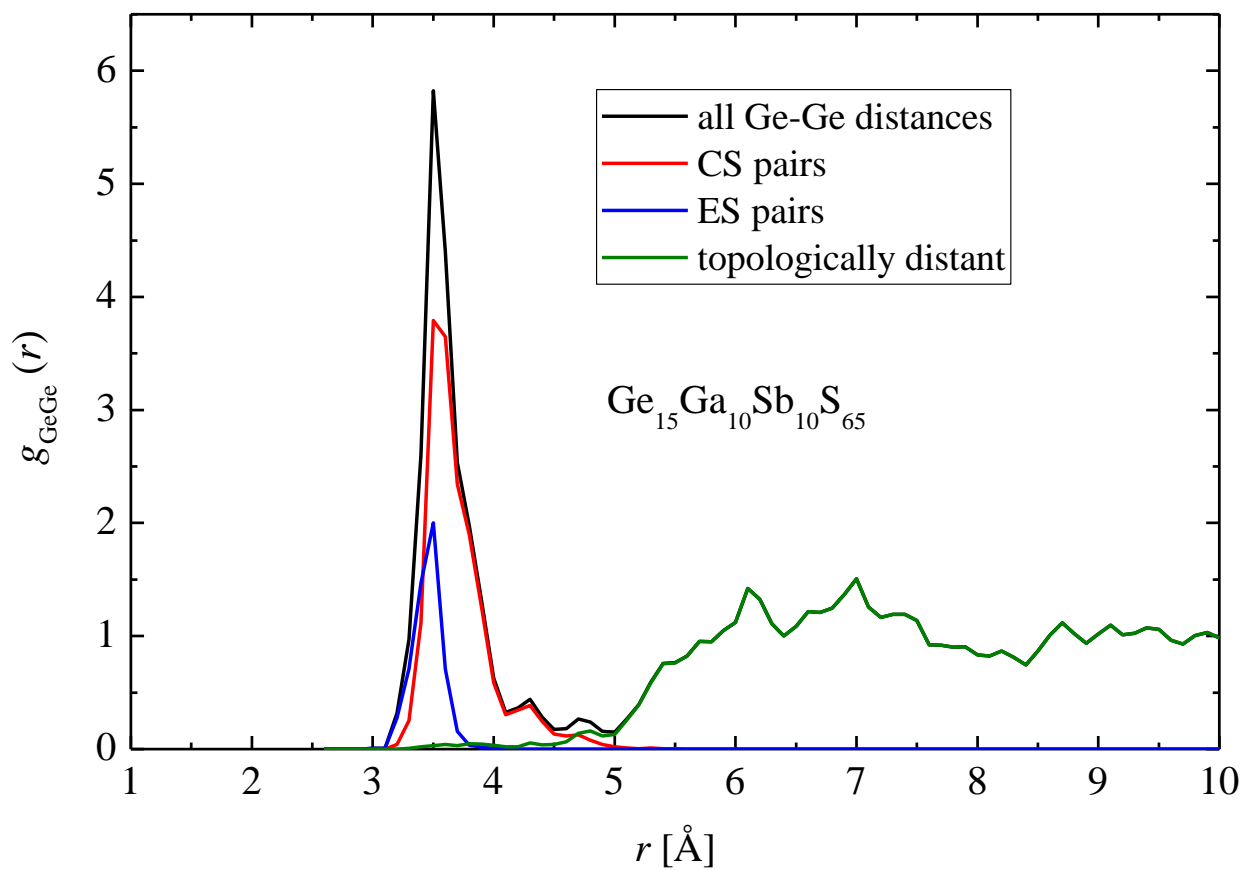

**Figure 6.** Decomposition of $g_{GeGe}(r)$ of $Ge_{15}Ga_{10}Sb_{10}S_{65}$ to contributions from corner shared (CS), edge shared (ES) tetrahedra and topologically distant Ge-Ge pairs.

**Figure 7** Neutron weighted partial structure factors of the $Ge_{15}Ga_{10}Sb_{10}S_{65}$ glass obtained by RMC simulation. Ga-X (X=Ga, Sb) partial structure factors are rather similar to the corresponding Ge-X ones, therefore they are not shown for clarity.